\PassOptionsToPackage{pdftex}{graphicx} % for arXiv process
\documentclass[preprint,11pt]{elsarticle}
\usepackage{lineno,hyperref}  % for arXiv process

\usepackage[pdftex]{graphicx}  % for arXiv process
\modulolinenumbers[100]  % for arXiv process

\usepackage[top=30truemm,bottom=30truemm,left=30truemm,right=30truemm]{geometry}
\journal{arXiv}

\begin{document}

\begin{frontmatter}

%% use the tnoteref command within \title for footnotes;
%% use the tnotetext command for the associated footnote;
%% use the fnref command within \author or \address for footnotes;
%% use the fntext command for the associated footnote;
%% use the corref command within \author for corresponding author footnotes;
%% use the cortext command for the associated footnote;
%% use the ead command for the email address,
%% and the form \ead[url] for the home page:
%%
%% \title{Title\tnoteref{label1}}
%% \tnotetext[label1]{}
%% \author{Name\corref{cor1}\fnref{label2}}
%% \ead{email address}
%% \ead[url]{home page}
%% \fntext[label2]{}
%% \cortext[cor1]{}
%% \address{Address\fnref{label3}}
%% \fntext[label3]{}

%\title{Separation of gamma-rays from neutrons with a CsI(Tl) scintillator 
%       using pulse shape discrimination method for gamma-ray spectroscopy 
%       under neutron rich field}
\title{{\bf Separation of gamma-ray and neutron events \\ with CsI(Tl) pulse shape analysis}}
  
%% use optional labels to link authors explicitly to addresses:
%% \author[label1,label2]{<author name>}
%% \address[label1]{<address>}
%% \address[label2]{<address>}

\author[kyoto]{Y. Ashida\corref{cor1}}
\ead{assy@scphys.kyoto-u.ac.jp}

\author[okayama]{H. Nagata}
\author[okayama]{Y. Koshio}
\author[kyoto]{T. Nakaya}
\author[kyoto]{R. Wendell}

\cortext[cor1]{Corresponding author}
\address[kyoto]{Department of Physics, Kyoto University, Kyoto 606-8502, Japan}
\address[okayama]{Department of Natural Science and Technology, Okayama University, 
                  Okayama 700-8530, Japan}

%///////////////////////////////////////////////////////////////////////////////////////
%  Abstract
%///////////////////////////////////////////////////////////////////////////////////////
\begin{abstract}
Fast neutrons are a large background to measurements of gamma-rays emitted
from excited nuclei, such that detectors that can efficiently distinguish
between the two are essential.
In this paper we describe the separation of gamma-rays from neutrons with the pulse shape
information of the CsI(Tl) scintillator, using a fast neutron beam and several gamma-ray sources.
We find that a figure of merit optimized for this separation
takes on large and stable values (nearly 4) between 5 and 10~MeV of electron equivalent deposited energy,
the region of most interest to the study of nuclear de-excitation gamma-rays.
Accordingly this work demonstrates the ability of CsI(Tl) scintillators
to reject neutron backgrounds to gamma-ray measurements at these energies.
\end{abstract}

%\begin{keyword}

%% keywords here, in the form: keyword \sep keyword

%% MSC codes here, in the form: \MSC code \sep code
%% or \MSC[2008] code \sep code (2000 is the default)

%\end{keyword}

\end{frontmatter}

%%
%% Start line numbering here if you want
%%
%\linenumbers

%///////////////////////////////////////////////////////////////////////////////////////
%  Sections
%///////////////////////////////////////////////////////////////////////////////////////
%\clearpage
%///////////////////////////////////////////////////////////////////////////////////////
%  Introduction 
%///////////////////////////////////////////////////////////////////////////////////////
\section{Introduction} 
\label{sec:intro}

%///////////////////////////////////////////////////////////////////////////////////////
%\subsection{Neutron background for gamma ray measurements} 

Measurements of the gamma-rays produced when excited nuclei relax to lower energy levels 
provide information on nuclear structure.
These gamma-rays are typically below 20~MeV in energy and 
are sensitive probes of not only the interaction between the exciting 
particles and target nuclei, but also the internal nucleon kinematics of the target itself. 
However, observation of such de-excitation photons from the interactions of nuclei with hadronic 
projectiles, such as protons, neutrons, and ions, is often subject to hadronic backgrounds.
Proton and ion backgrounds can be removed through a variety of methods, such as 
using anti-coincidence counting of the primary gamma-ray detector and a 
dedicated charged particle detector placed between it and the target.
Fast neutron backgrounds, on the other hand, are more problematic. 
Placing thick shielding made of heavy materials upstream of the photon detector is 
a common method of reducing these backgrounds, whose main disadvantage is the attenuation of 
gamma-rays as well.
For this reason the ideal photon detector should be able to distinguish gamma-ray events 
from neutron-induced events. 

To achieve this ideal, pulse shape discrimination techniques are applied to detectors 
whose signal shape changes in response to excitation from different particles.
Organic liquid scintillators, such as BC-501A, are often used for this purpose~\cite{leo,knoll},
as well as for neutron detection in general~\cite{iwamoto}. 
However such scintillators are not well suited to precision gamma-ray measurements 
since photons primarily interact with organic materials via Compton scattering 
leading to energy loss and subsequently poor energy resolution and reduced detection efficiency.
Germanium detectors, on the other hand, are often used for gamma-ray spectroscopy, but 
are not sufficiently radiation hard to handle measurements in a neutron environment.
Further, these detectors are not ideal for gamma-ray multiplicity measurements, 
which require large acceptances to cover all target nuclei, due to their high cost 
and handling difficulty.
Inorganic scintillators provide a viable alternative for both gamma-ray spectroscopy and 
multiplicity measurements.
Among such scintillators, CsI(Tl) is promising photon detector material, due to its 
high density ($4.51 \ {\rm g/cm^3}$), high light yield (1.7 times higher than 
NaI(Tl)), and physical scalability.
Further, CsI(Tl) is known to have the ability to separate 
electron and nuclear recoils~\cite{wu}, 
to separate gamma-rays from alpha particles~\cite{dcunha},
and to distinguish among several types of ions~\cite{alarja,alderighi}.
Despite these successes there have been few studies focused on separation of gamma-rays and neutrons, 
which is important for addressing the ability of CsI(Tl) detectors to make measurements 
in the presence of large neutron backgrounds.
For example, when fast neutrons with O(10)~MeV kinetic energies are incident on a water target 
surrounded by detectors to measure the subsequent gamma-ray production, such as in Ref.~\cite{huang}, 
such ability gives benefits to the measurements. 
Such measurements are of further interest to constrain O(1)~MeV gamma-ray backgrounds 
from neutron-oxygen reactions in the measurement of neutral current neutrino-oxygen 
cross section at the T2K experiment~\cite{t2kncqe}.

In this paper we evaluate the separation of gamma-rays from neutrons with pulse shape discrimination 
(PSD) techniques in CsI(Tl) scintillators using the neutron time-of-flight beam line at Tohoku University's 
Cyclotron and Radioisotope Center (CYRIC)~\cite{orihara}.
%Fast neutrons with O(10)~MeV kinetic energies are incident on a water target 
%surrounded by detectors as shown Figure~\ref{fig:ncgammaexp} 
%to measure the subsequent gamma-ray production~\cite{huang}.
%Such measurements are of further interest to constrain O(1)~MeV gamma backgrounds 
%from neutron-oxygen reactions in the measurement of the neutral current neutrino-oxygen 
%cross section at the T2K experiment~\cite{t2kexp,t2kncqe}.
The paper is organized as follows.
Section~\ref{sec:psd} discusses PSD techniques in general, 
and Section~\ref{sec:experiment} describes the experiment itself. 
The PSD performance of the CsI(Tl) scintillators is evaluated in  
Section~\ref{sec:analysis} before concluding in Section~\ref{sec:conlusion}

% \begin{figure}[htbp]
%  \begin{center}
%   \includegraphics[clip,width=11.0cm]{./figtbl/ncgammaexp.pdf}
%  \end{center}
%  \vspace{-110truept}
%  \caption{Schematic draw of the gamma ray production measurement. 
%           A quasi-mono energetic neutron beam is incident on a water target 
%	   and gamma rays from neutron-oxygen reactions are measured with gamma ray 
%	   detectors around the target. In order to measure the multiplicity
%           of gamma rays, large acceptance detector is required.}
%  \label{fig:ncgammaexp}
% \end{figure}

%///////////////////////////////////////////////////////////////////////////////////////
\section{Pulse shape discrimination technique} 
\label{sec:psd}

Pulse shape discrimination methods achieve particle identification based on  
information carried in the detector's output waveform.
The waveform for a class of scintillation materials can be 
described by two decaying components as
 
  \begin{eqnarray}
   I(t) = I_{\rm fast} \cdot \exp\left({-\frac{t}{\tau_{\rm fast}}}\right) 
        + I_{\rm slow} \cdot \exp\left({-\frac{t}{\tau_{\rm slow}}}\right),  
  \label{eq:scintiyield}
  \end{eqnarray}
  \vspace{1truept}
 
\noindent
where $I(t)$ is the light yield as a function of time $t$. 
The value of the $I_{\rm fast}$ ($I_{\rm slow}$) constant and $\tau_{\rm fast}$ ($\tau_{\rm slow}$)
refer to the light yield and 
decay time constant of the fast (slow) component, where $\tau_{\rm fast} < \tau_{\rm slow}$.
It should be noted that decay constants are specific to the scintillation material.
When the ratio of $I_{\rm fast}$ to $I_{\rm slow}$ depends on the rate of energy loss over 
the travel distance inside the material ($-dE/dx$), particle identification is possible.

The scintillation mechanism of inorganic materials is characteristic of 
their electronic band structures, as detailed in Refs.~\cite{leo,knoll}.
When a particle deposits energy inside an inorganic material,
electrons are excited from the valence to the conduction band, producing electron-hole 
pairs in the scintillator.
Though the promoted electrons can rapidly de-excite via the emission of photons,
electron-hole bound states known as ``excitons'' can also be formed.
Such states similarly decay via photon emission but on shorter timescales 
than de-excitation from the conduction band. 
The probability of creating an exciton increases with the number of created 
electron-hole pairs, which is itself proportional to the deposited energy.
Accordingly, $I_{\rm fast}$ will increase relative to $I_{\rm slow}$ for 
larger energy depositions (larger $|-dE/dx|$).
Note that this is the opposite of the behavior of organic scintillators, which 
scintillate via entirely different mechanisms.
For this reason the ratio of $I_{\rm fast}$ and $I_{\rm slow}$ depends on the incident particle's 
$|-dE/dx|$, and this information can be used for particle identification.
Typical values of $\tau_{\rm fast}$ and $\tau_{\rm slow}$ for CsI(Tl) are 680~ns and 3340~ns, respectively~\cite{knoll}.

%///////////////////////////////////////////////////////////////////////////////////////

%%For example, the neutron reaction with nuclei above 20 MeV of kinetic energy is not studied 
%%experimentally well. 
%%The typical energy of nuclear de-excitation gamma-rays is below 10 MeV, and 20 MeV at largest. 
%%This kind of detector is called as phoswich type detector, and for instance, the plastic 
%%scintillator is used as sub-detector. 

%\clearpage
%///////////////////////////////////////////////////////////////////////////////////////
%  Experimental details 
%///////////////////////////////////////////////////////////////////////////////////////
\section{PSD capability test} 
\label{sec:experiment}

%///////////////////////////////////////////////////////////////////////////////////////
%%\subsection{Experimental details} 

In order to study the PSD capabilities of CsI(Tl) scintillators we have performed a 
set of experiments at CYRIC~\cite{orihara} using both neutron and gamma-ray beams.
We have investigated the separation performance at energies between 3 and 18 MeV. 

\begin{figure}[htbp]
  \begin{center}
   \includegraphics[clip,width=12.0cm]{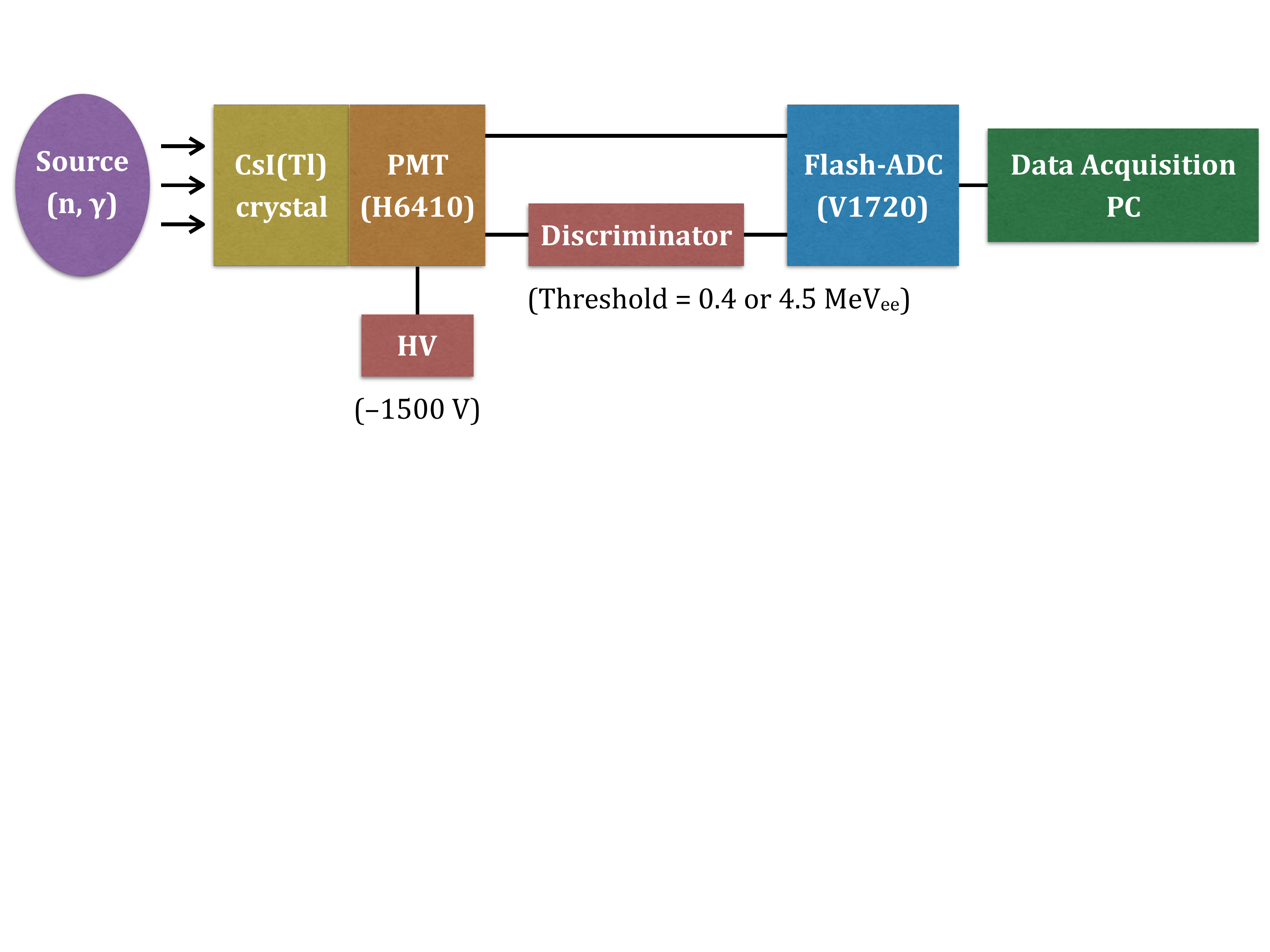}
  \end{center}
  \vspace{-150truept}
  \caption{Schematic diagram of the data-taking system for the CsI(Tl) PSD test. 
           Gamma-rays and fast neutrons are incident on the detector from the left.}
  \label{fig:readout}
 \end{figure}

The experiment uses an ${\rm OKEN^{\textregistered}}$ CsI(Tl) crystal of dimension 
$35 \times 35 \times 35 \ {\rm mm^3}$ optically coupled to
a ${\rm HAMAMATSU^{\textregistered}}$ H6410/R329-02 photomultiplier tube (PMT). 
The thallium doping amount to the crystal is 0.02\%.
A negative bias of $- 1500$ V is applied to the PMT using a REPIC HV module, 
and a CAEN V1720 Flash-ADC (12 bit, 250 MS/s) module is used for the waveform read out. 
The module receives discriminated signals from the PMT and is operated in the self-triggering mode,
as shown in Figure~\ref{fig:readout}.

 \vspace{-8truept}  % only for processing on arXiv

\begin{figure}[htbp]
  \begin{center}
   \includegraphics[clip,width=14.0cm]{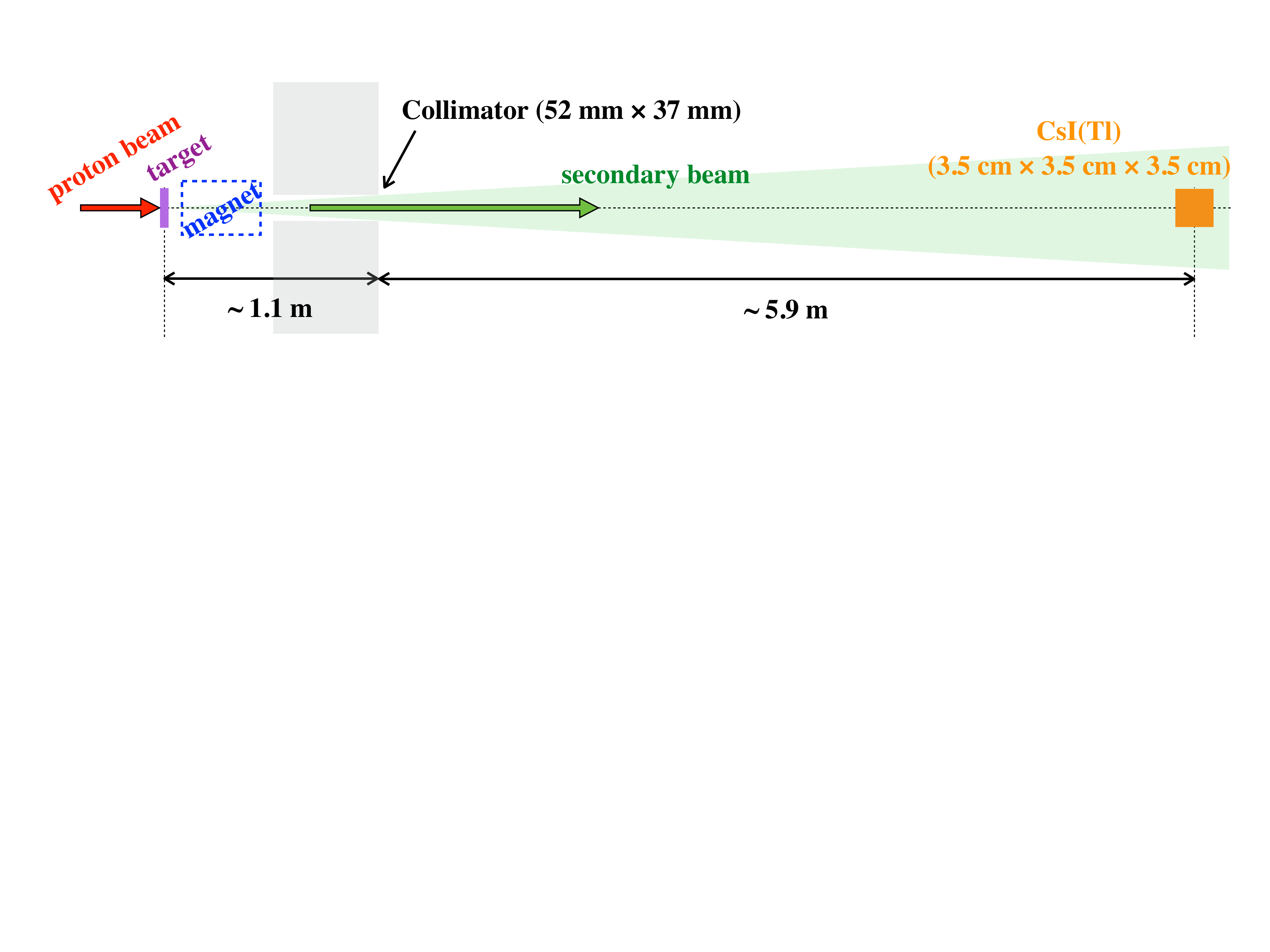}
  \end{center}
  \vspace{-195truept}
  \caption{Schematic view of the experiment. Secondary beams are made using 70 MeV 
           protons incident on either a copper or lithium target, to produce 
           beams enriched in gamma-rays or neutrons, respectively.
	   The CsI(Tl) crystal is located about 6~m downstream of the collimator.}
  \label{fig:setup}
 \end{figure}

At the CYRIC facility protons are accelerated by a cyclotron and injected on selected targets to produce secondary beams.
In this experiment two different target materials were used, a copper target for producing a gamma-ray-enriched 
beam and a lithium target for a neutron-rich beam via the ${\rm ^{7}Li(p,n)^{7}Be}$ reaction.
The proton kinetic energy was selected to be 70 MeV and resulted in neutrons 
with kinetic energies of several tens of MeV.  
The primary proton beam current was set to 1~nA to ensure negligible event pileup.
The beam provides bunches with a 50~ns period and 
is collimated using a collimator made of iron and concrete with $52 \times 37 \ {\rm mm^2}$  cross section 
and 0.5~m thickness situated just behind the target.
During beam operations the CsI(Tl) detector is placed 6 m away from the collimator on the beam axis.
Charged particles produced before the collimator are removed by the bending magnet, 
therefore only neutral particles (photons and neutrons at this energy region) enter the beam line.
Figure~\ref{fig:setup} shows a schematic view of the experiment.
The detector was calibrated using a ${\rm ^{60}Co}$ source as well as environmental radioisotopes, such as ${\rm ^{40}K}$.

%///////////////////////////////////////////////////////////////////////////////////////
%%\subsection{Data sample} 

%\newpage
The data from our measurements are categorized into three gamma-ray-rich samples,
one taken with only environmental backgrounds (sample A), 
one taken with the ${\rm ^{60}Co}$ (sample B),  
and one under a beam run with the copper target (sample C), 
and one neutron-rich sample taken from beam running with the lithium target (sample D).
The data set information is summarized in Table~\ref{tab:dataset}.
An energy threshold of 0.4 ${\rm MeV_{ee}}$ (MeV electron equivalent) 
was applied to all samples. 
In addition, a separate measurement for sample D using the lithium target and a higher threshold, 4.5 ${\rm MeV_{ee}}$,
was taken to study neutrons with minimal gamma-ray contamination.
In Table~\ref{tab:dataset}, low (high) corresponds to 0.4 (4.5) ${\rm MeV_{ee}}$.

 \begin{table}[htbp]
  \begin{center}
  \caption{Data sample information for the CsI(Tl) PSD test.}
  \label{tab:dataset}
  \vspace{2truept}
   \begin{tabular}{l | l | c} \hline \hline
    Sample        & Source & Number of events \\ \hline
    A (gamma-ray) & Environmental background (${\rm ^{40}K}$, etc) & 10605 \\
    B (gamma-ray) & ${\rm ^{60}Co}$ source & 9992 \\
    C (gamma-ray) & Beam with the copper target & 7905 \\
    D (neutron)   & Beam with the lithium target & 49504 (low) $+$ 29989 (high) \\ \hline \hline
   \end{tabular}
  \end{center}
 \end{table}
%%
%%

%///////////////////////////////////////////////////////////////////////////////////////

%\clearpage
%///////////////////////////////////////////////////////////////////////////////////////
%  Analysis result 
%///////////////////////////////////////////////////////////////////////////////////////
\section{CsI(Tl) PSD performance evaluation}  
\label{sec:analysis}

%///////////////////////////////////////////////////////////////////////////////////////
\subsection{Waveform analysis}

 \begin{figure}[h]
  % 1st figure
  \begin{minipage}{0.5\hsize}
   \begin{center}
    \includegraphics[clip,width=7.7cm]{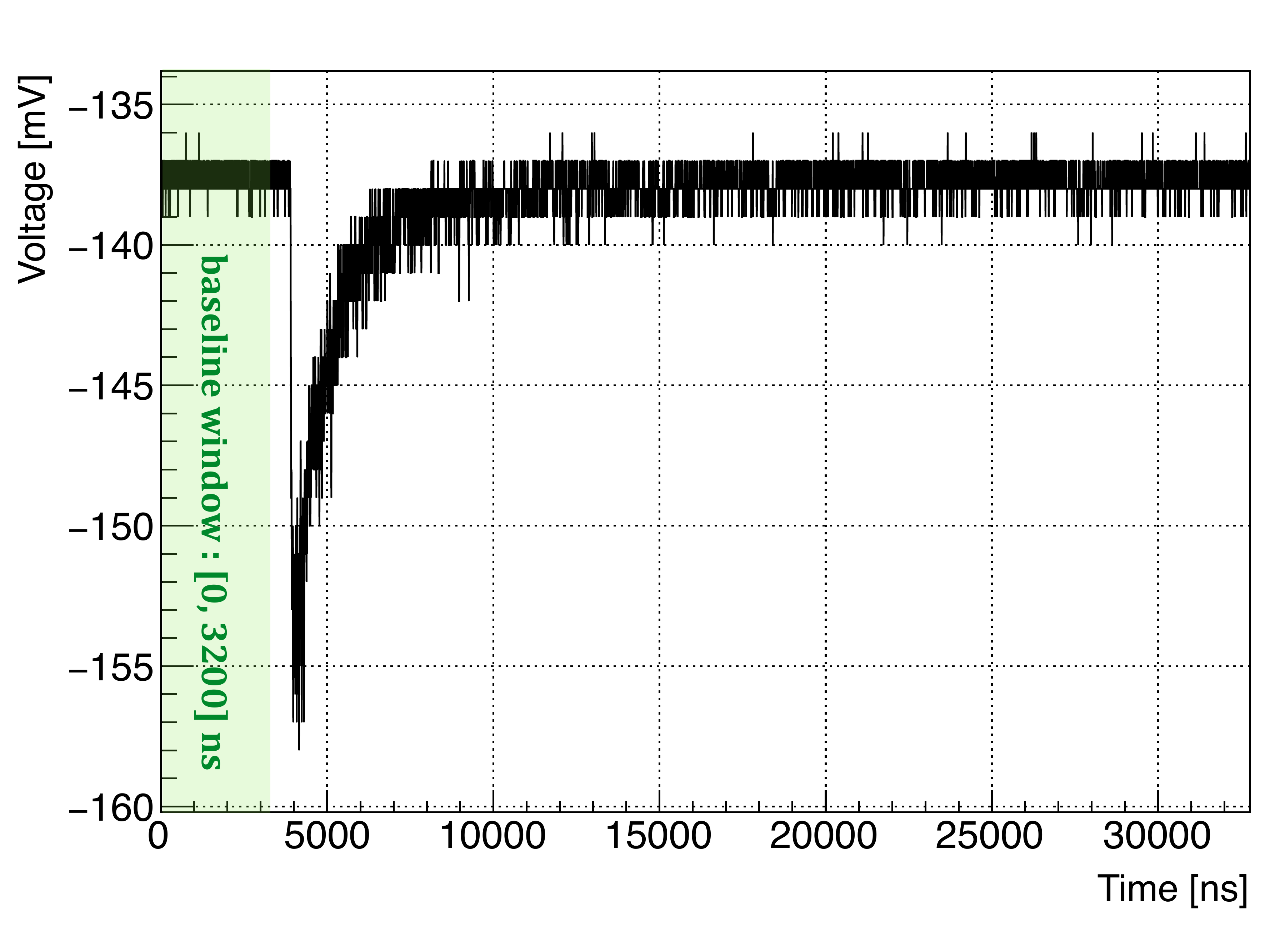}
   \end{center}
  \end{minipage}
  % 2nd figure
  \begin{minipage}{0.5\hsize}
   \begin{center}
    \includegraphics[clip,width=7.7cm]{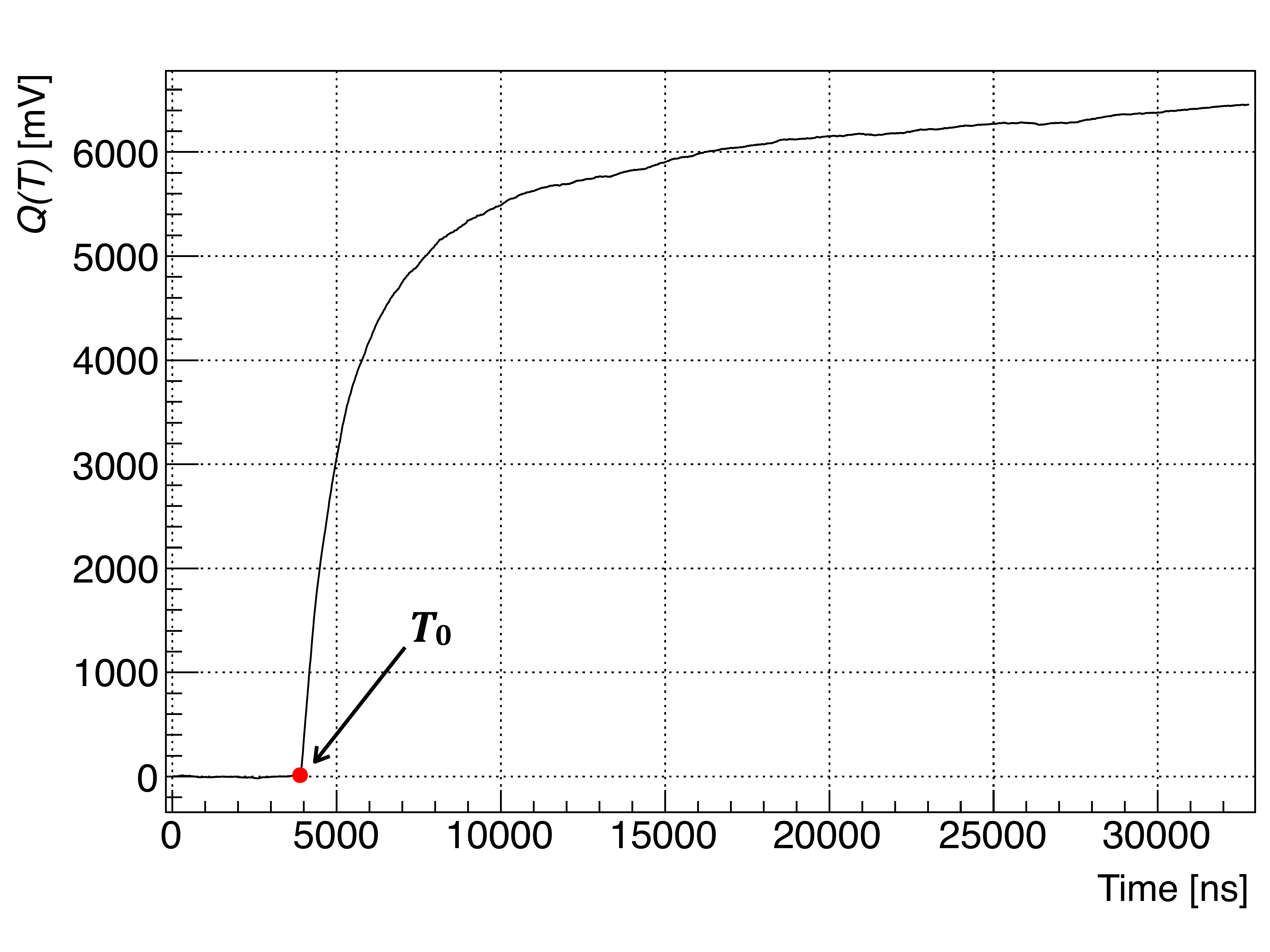}
   \end{center}
  \end{minipage}
  \vspace{-10truept}
  \caption{Left: An example CsI(Tl) waveform. 
           The green hatched area is the baseline calculation window.
           Right: An example $Q(T)$ distribution obtained from the waveform in the left panel.
	   The analysis start point $T_0$ is shown by the red marked circle.} 
  \label{fig:waveplot}
 \end{figure}

Waveform data in the region 4096~ns prior to and up to 28672~ns after the event trigger were saved using 8192 samples.
An example waveform is shown in the left panel of Figure~\ref{fig:waveplot}.
For each trigger the baseline, $\mu$, for the event is calculated as the average amplitude between 0 and 3200 ns.
Then the cumulative charge $Q(T)$ is calculated as the integral 
of the waveform voltage taken between time 0 and time $T$ after subtracting the baseline,

 \begin{eqnarray}
  Q(T) = \int_{0}^{T} \left(|{\rm Voltage}(T')| - |\mu|\right) dT'.
 \label{eq:scintiyield}
 \end{eqnarray}
 \vspace{1truept}

\noindent
The right panel of Figure~\ref{fig:waveplot} shows $Q(T)$ for the waveform in the left panel.
Here a red circle marks the analysis start point, referred to as $T_0$ from here onward.
The point $T_0$ is determined as the first point where $Q(T)$ increases continuously for more than 400 ns.
%%This criteria is tested to be good with checking by eye for randomly picked out events
%%for every data sample. 

\begin{figure}[htbp]
  \begin{center}
   \includegraphics[clip,width=9.5cm]{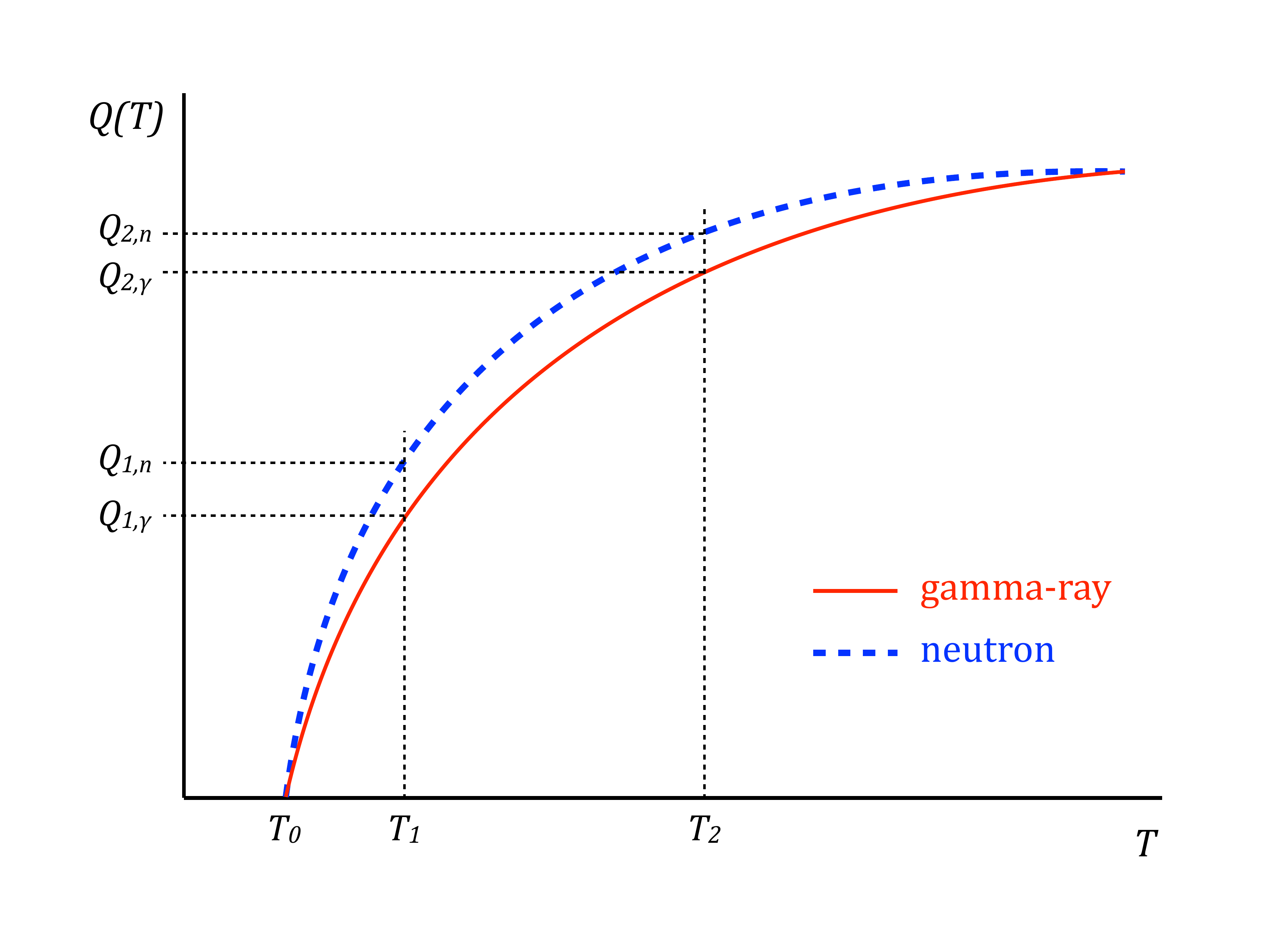}
  \end{center}
  \vspace{-20truept}
  \caption{Schematic drawing of $Q(T)$ distributions for gamma-ray- and neutron-induced events.
           The plots are normalized by event total energy.
	   $T_0$ is the time at the analysis start point.}
  \label{fig:ngqtplot}
 \end{figure}

%\newpage
The shape of the $Q(T)$ distribution is characteristic of the scintillation constants expressed in the raw waveform.
As explained in the previous section, the energy deposition of neutrons and gamma-rays differ
and accordingly their $Q(T)$ distributions will differ as well,
even after accounting for differences in the total energy deposited, as shown in Figure~\ref{fig:ngqtplot}.
Particle identification can be performed using these differences.
Here, total energy is the value of $Q(T)$ when $T\rightarrow\infty$.
Since the probability of waveform contamination from event pileup increases
at large $T$ and because $Q(T)$ 
may not saturate within the DAQ window for large signal, we opt to study the 
PSD performance of the scintillator at two values of $T$. 
We label these values $T_{1}$ and $T_{2}$, with $T_{1} < T_{2}$, and their respective 
cumulative charges are then $Q_{1}$ and $Q_{2}$.
Particle discrimination can be performed using the ratio of these values, $Q_{1}/Q_{2}$,
since this quantity is in general expected to differ for neutron and gamma-ray waveforms, 
as shown in Figure~\ref{fig:ngqtplot}.

%%Here there is no regulation on the selection of $T_{1}$ and $T_{2}$.
%%When the ratio of $Q_{1}$ and $Q_{2}$ is made for gamma-ray (red marked) 
%%and neutron (blue marked) events, their values should be different 
%%($Q_{1,R}/Q_{2,R} \neq Q_{1,B}/Q_{2,B}$), then separation can be done.

%%
%%

The energy of each event is defined to be the integrated charge between 
$T_{0}$ and $T_{2}$ such that, $T_2 - T_0 = 4 \ {\rm \mu s}$.
This serves to define the value of $T_{2}$. 
The value of $T_1$ is optimized in the analysis and will be discussed below.
%%The region between $T_1$, which is optimized in the analysis as discussed below, and 
%%$T_{2}$ is termed as the 
As mentioned above, the detector's energy calibration is performed using 
the observed photoelectron absorption peaks from the decays of ${\rm ^{60}Co}$ and ${\rm ^{40}K}$. 
For neutron-induced events the visible energy value is expected to be shifted 
from the true one due to quenching in the scintillator.
For this reason we use electron equivalent energies, defined as the energy of an electron 
that would produce the observed visible energy, as the energy estimator.
Note that it is not necessary to measure the true neutron energy deposition 
for the gamma-ray measurements. 
%Figure \ref{fig:calibplot} shows the result of the calibration. Red line shows the fit function 
%and the fit looks to be done well. From this, energy calibration constants are obtained. 
%
% \begin{figure}[htbp]
%  \begin{center}
%   \includegraphics[clip,width=8.0cm]{./figtbl/calibplot.pdf}
%  \end{center}
%  \vspace{-20truept}
%  \caption{Result of the energy calibration. Gamma-ray peaks from ${\rm ^{60}Co}$ 
%           and ${\rm ^{40}K}$ are used. Calibration constants are obtained by 
%	   linear fitting, and the fit looks good.} 
%  \label{fig:calibplot}
% \end{figure}

%///////////////////////////////////////////////////////////////////////////////////////
\subsection{Separation performance}

The CsI(Tl) scintillator's PSD performance is evaluated using the parameter

 \begin{eqnarray}
  {\rm PSD \ parameter} = 
  %\frac{Q(T_2 = T_0 + 4 \ {\rm \mu s}) - Q(T_1)}{Q(T_2 = T_0 + 4 \ {\rm \mu s})}, 
  \frac{Q(T_2) - Q(T_1)}{Q(T_2)}. 
 \label{eq:psdparameter}
 \end{eqnarray}
 \vspace{1truept}

\begin{figure}[h]
  % 1st figure
  \begin{minipage}{0.5\hsize}
   \begin{center}
    \includegraphics[clip,width=7.75cm]{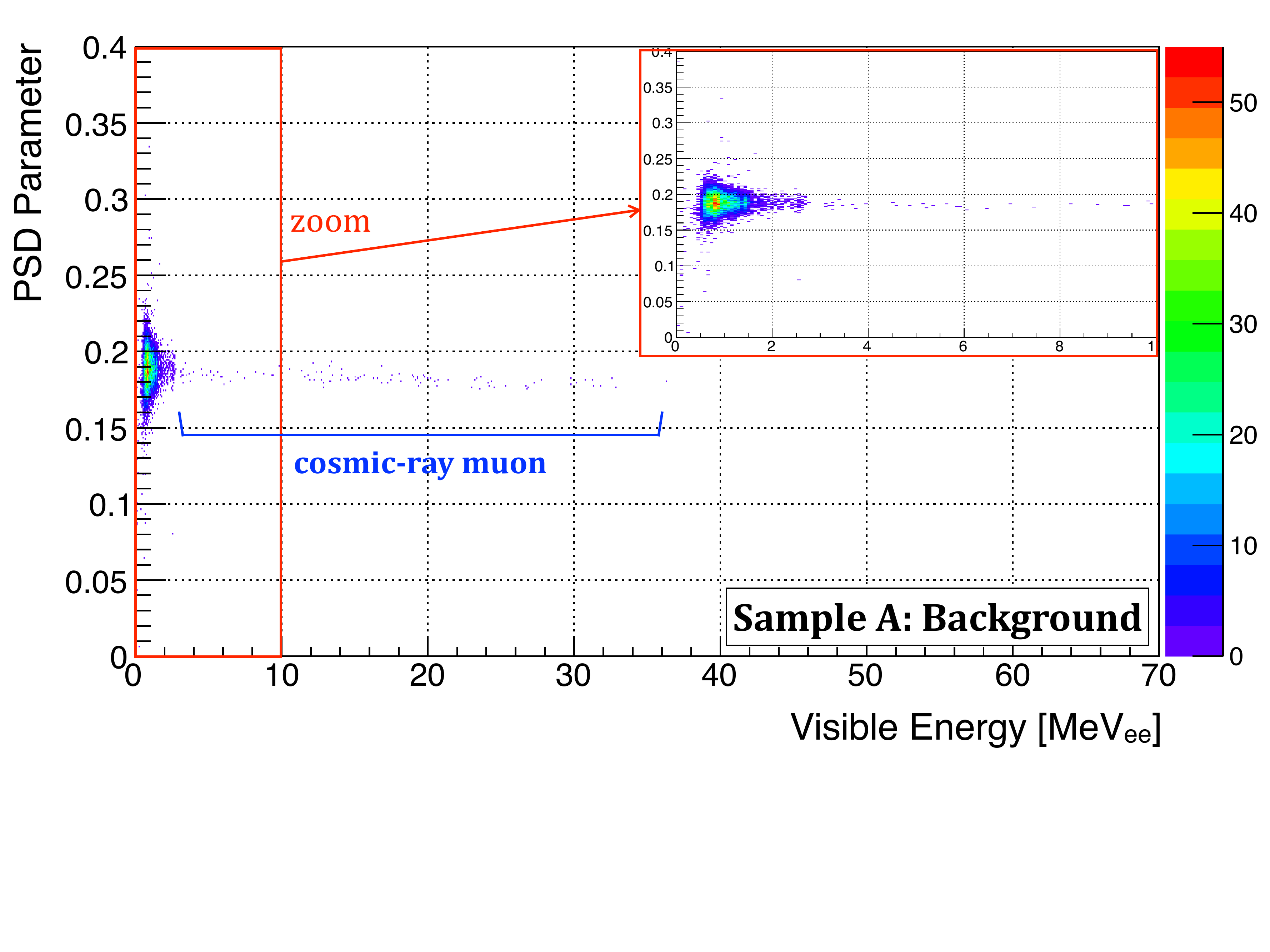}
   \end{center}
  \end{minipage}
  % 2nd figure
  \begin{minipage}{0.5\hsize}
   \begin{center}
    \includegraphics[clip,width=7.75cm]{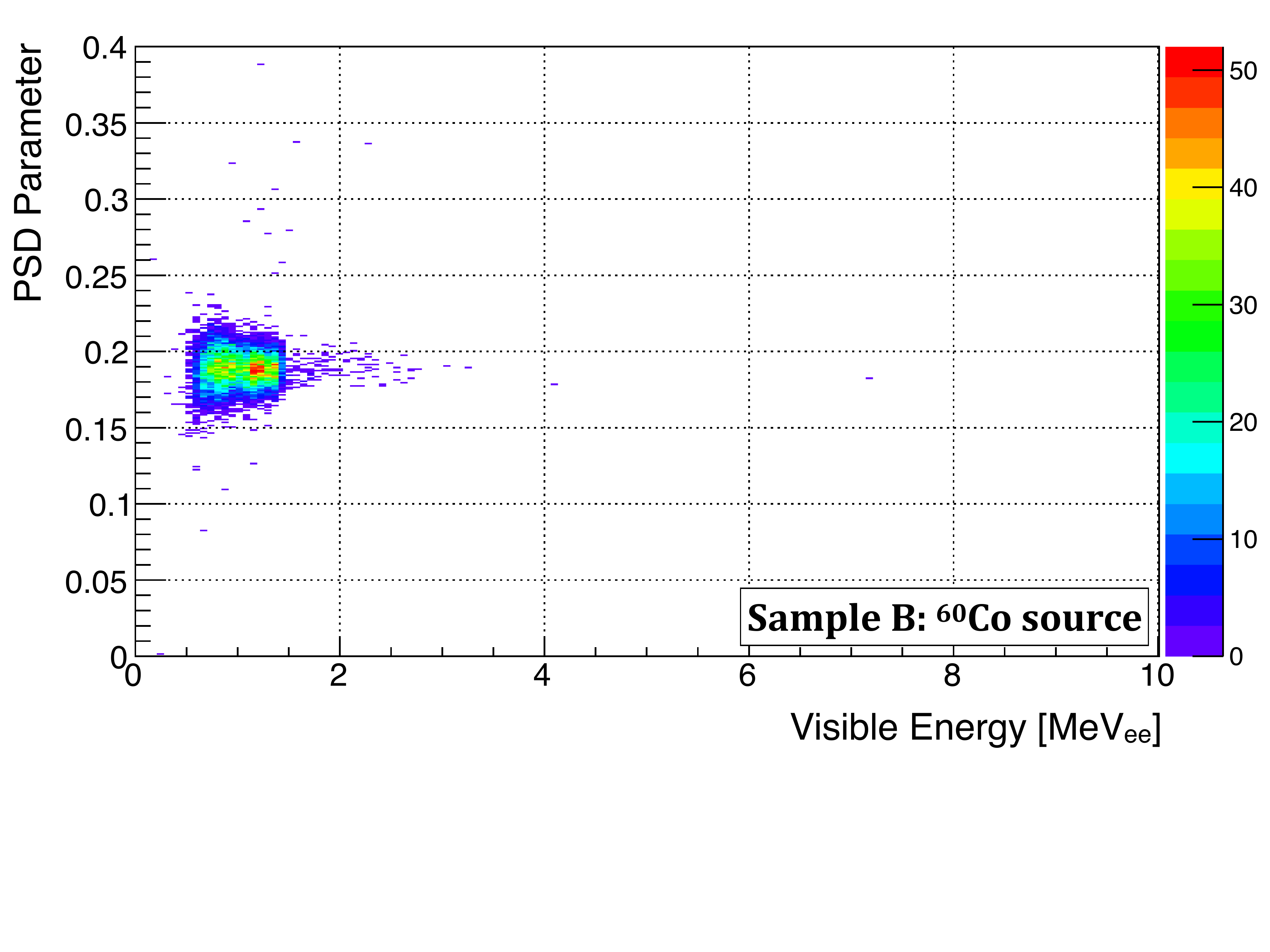}
   \end{center}
  \end{minipage}
  % 3rd figure
  \begin{minipage}{0.5\hsize}
  \vspace{-30truept}
   \begin{center}
    \includegraphics[clip,width=7.75cm]{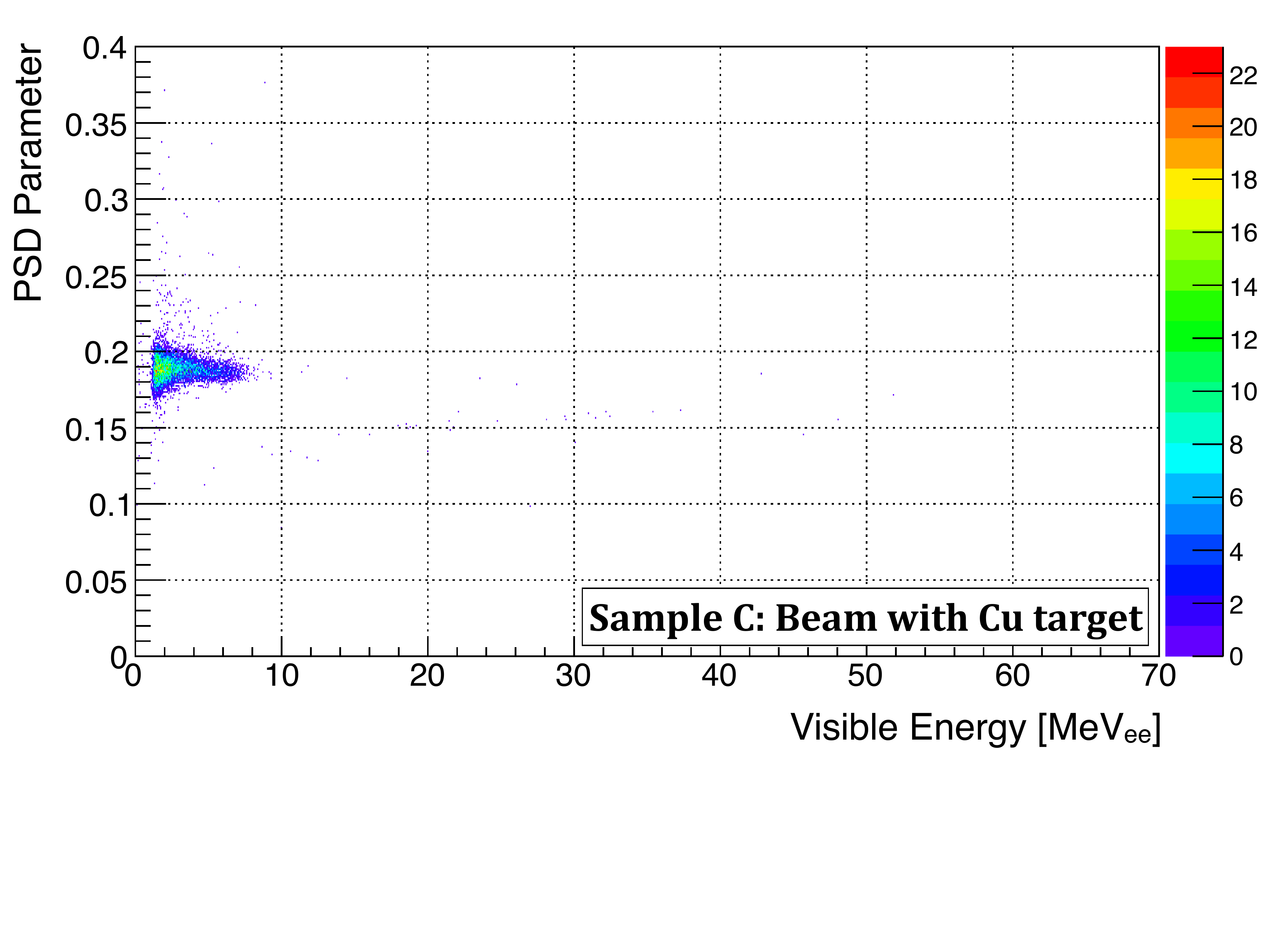}
   \end{center}
  \end{minipage}
  % 4th figure
  \begin{minipage}{0.5\hsize}
  \vspace{-30truept}
   \begin{center}
    \includegraphics[clip,width=7.75cm]{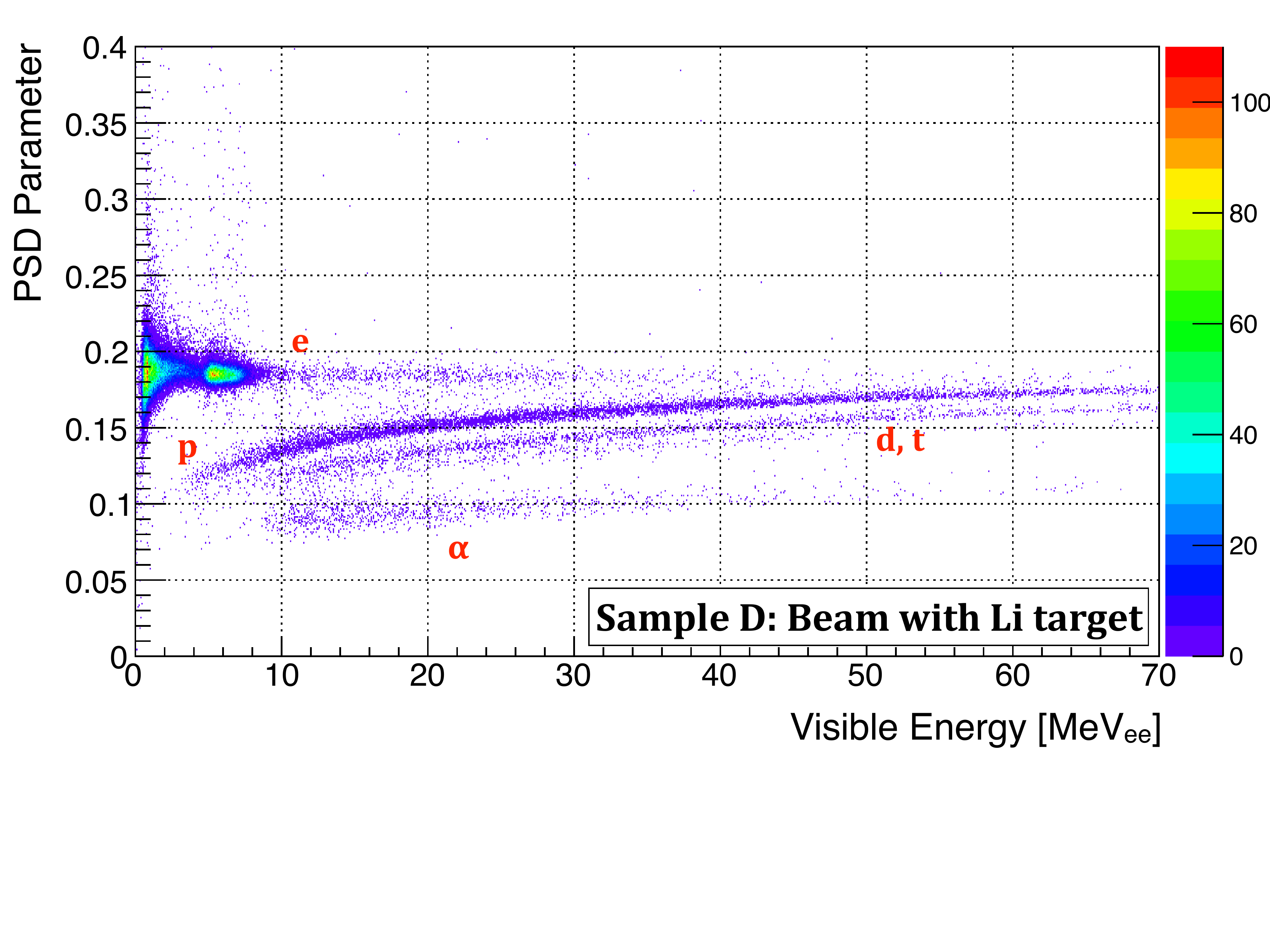}
   \end{center}
  \end{minipage}
  \vspace{-35truept}
  \caption{Distribution of the PSD parameter as a function of visible energy, with $T_1$ set to $T_0 + 2 \ {\rm \mu s}$.
           Data from the gamma-ray-enriched samples A, B, and C, appear in the top left, top right, and bottom left 
           panels, respectively. The bottom right plot shows the neutron-rich sample D.}
  \label{fig:psdplots}
 \end{figure}

\noindent
This parameter expresses the ratio of the late-time charge deposition, also termed the ``tail'', 
to the total deposition and is expected to be larger for gamma-ray events relative to neutron events.  
Figure~\ref{fig:psdplots} shows the PSD parameter  as a function of visible energy for each of 
the data samples. 

\begin{figure}[htbp]
  \begin{center}
   \includegraphics[clip,width=9.5cm]{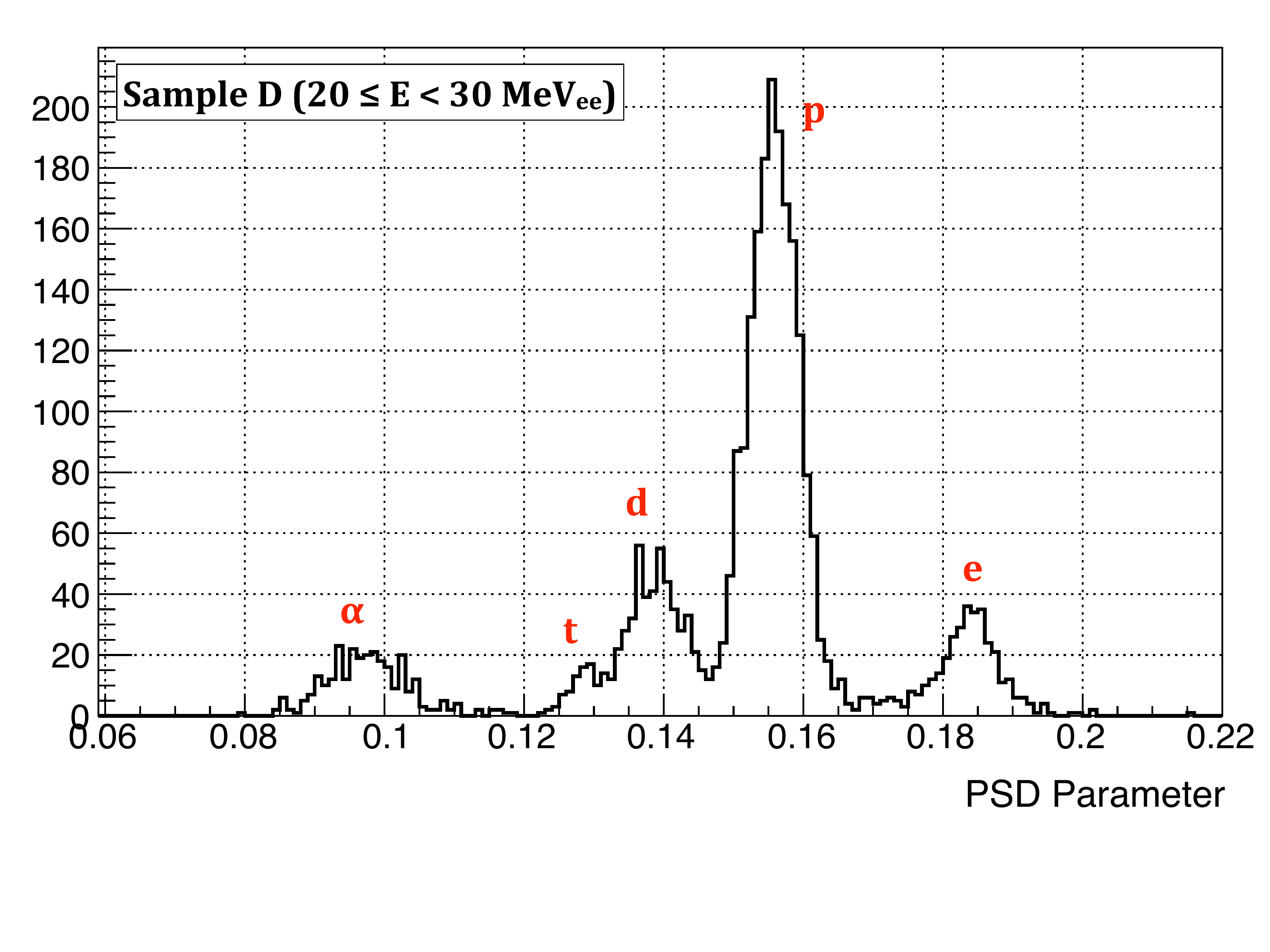}
  \end{center}
  \vspace{-35truept}
  \caption{PSD parameter distribution from sample D for events with visible energy  
           between 20 and 30 ${\rm MeV_{ee}}$. 
	   A clear separation among electron, proton, deuteron, triton, and alpha particle events is visible.}
  \label{fig:psd1dplot}
 \end{figure}

\noindent
Here, $T_1$ is set to $2 \ {\rm \mu s}$ from $T_0$. 
Samples A, B, and C, all of which are enriched in gamma-rays, show a concentrated 
collection of events with PSD parameters around values of 0.18. 
There are several events in the sample A data at energies
up to $\sim 35 \ {\rm MeV_{ee}}$ that are on the same PSD parameter line as the lower energy gamma-ray 
events.
These are considered to be contributions from cosmic-ray muons. 
In contrast, the neutron-enriched sample D has several distinct event populations
that are not seen in the other plots. 
These populations can be identified as protons, deuterons, tritons, and alpha particles (in descending 
order of PSD parameter), based on their expected $|-dE/dx|$  values.
These particles are caused by neutron reactions inside the detector, such as ${\rm (n,p)}$ and ${\rm (n,\alpha)}$. 
Events that lie at the same PSD parameter values as the gamma-ray samples, but at higher energies, are 
considered to be electron scatters produced by a variety of processes in the sample D data set.
The two clusters of events at lower energies at the same PSD parameter value 
are an artifact of merging data with separate thresholds in this sample.
Figure~\ref{fig:psd1dplot} shows the PSD parameter distribution 
for events with visible energy between 20 and 30 ${\rm MeV_{ee}}$, making the contribution 
from each particle species clearer.

\begin{figure}[htbp]
  \begin{center}
   \includegraphics[clip,width=9.5cm]{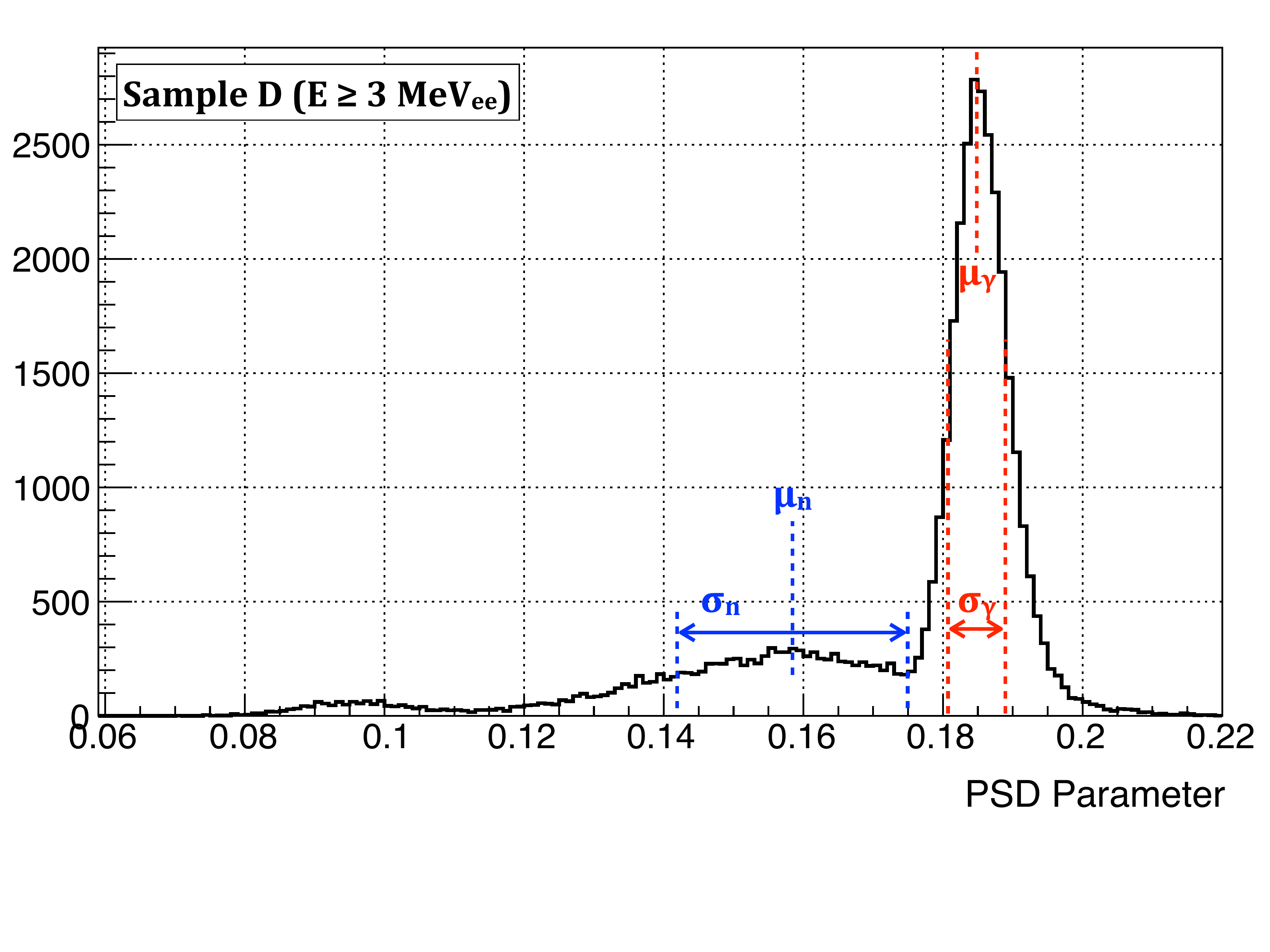}
  \end{center}
  \vspace{-35truept}
  \caption{Sample D's PSD parameter distribution for events with visible energy  
           larger than 3 ${\rm MeV_{ee}}$. The position of the neutron 
           and gamma-ray peaks are identified by the $\mu_{\gamma}$ and $\mu_{n}$ 
           labels. Their FWHM values are similarly identified with the $\sigma$ 
           labels.} 
  \label{fig:psdevaluate}
 \end{figure}

Figure~\ref{fig:psdevaluate} shows the same projection but for all events with visible energy larger than 3 ${\rm MeV_{ee}}$. 
In order to evaluate the PSD performance we adopt a conventional figure of merit (FOM)~\cite{soderstrom} as follows: 

 \begin{eqnarray}
  {\rm FOM} = \frac{|\mu_\gamma - \mu_n|}{\sigma_\gamma + \sigma_n}, 
 \label{eq:psdparameter}
 \end{eqnarray}
 \vspace{1truept}

\noindent
where $\mu_{\gamma}$ ($\mu_{n})$ refers to the peak position and $\sigma_{\gamma}$ ($\sigma_{n}$) the 
full-width-half-maximum (FWHM) of the gamma-ray (neutron) peak.  
Here the gamma-ray peak is taken to be the tallest and the neutron peak the closest 
peak thereto.

\begin{figure}[htbp]
  \begin{center}
   \includegraphics[clip,width=9.5cm]{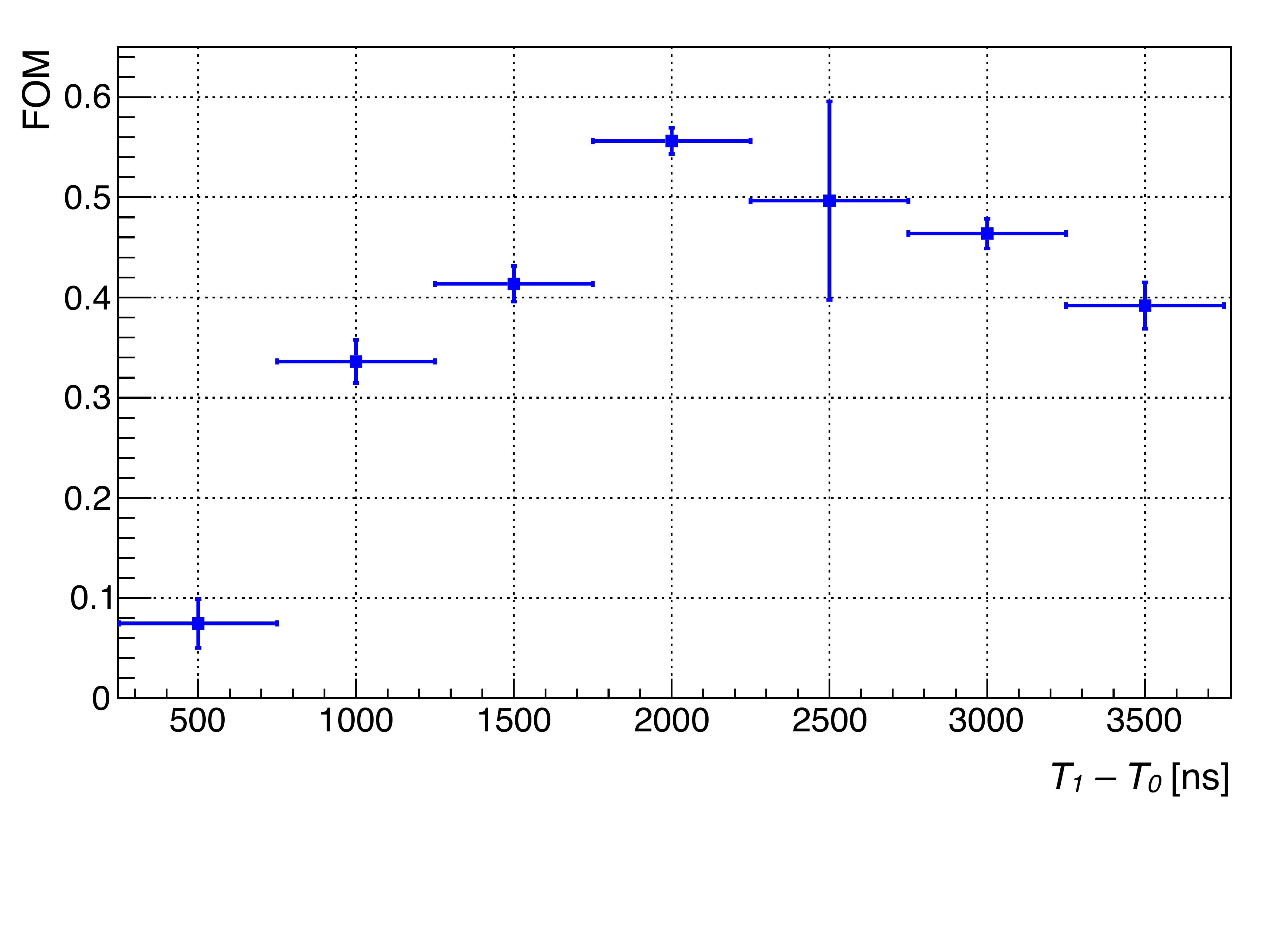}
  \end{center}
  \vspace{-40truept}
  \caption{Calculated figures of merit for different $T_1$ values. 
           The FOM becomes largest at $T_1 = T_0 + 2 \ {\rm \mu s}$.} 
  \label{fig:fomintegraltime}
 \end{figure}

\newpage
\noindent
The $T_1$ parameter is optimized by maximizing the FOM 
over different $T_1$ values, ranging from 500~ns to 
3500~ns after $T_{0}$ in steps of 500~ns.
The results are shown in Figure~\ref{fig:fomintegraltime},
where it can be seen that the FOM has its highest value at $T_1 = T_0 + 2 \ {\rm \mu s}$.

\begin{figure}[htbp]
  \begin{center}
   \includegraphics[clip,width=9.5cm]{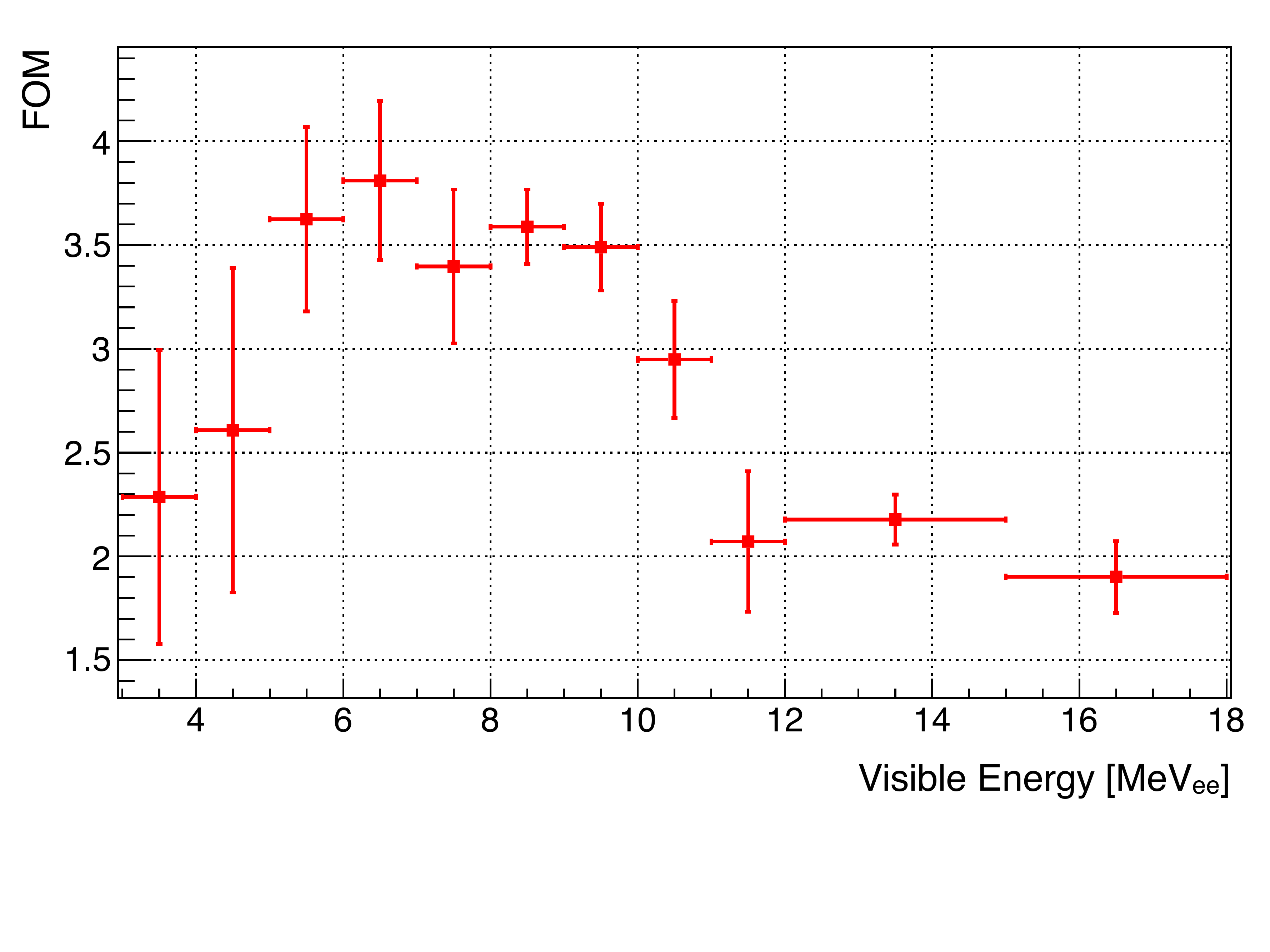}
  \end{center}
  \vspace{-40truept}
  \caption{Calculated figures of merit for different visible energy ranges. 
           The FOM takes on large but nearly constant values between 5 and 10~${\rm MeV_{ee}}$,
           the region of most interest for nuclear de-excitation gamma-ray measurements.} 
  \label{fig:fomenergyregion}
 \end{figure}

\noindent
Fixing $T_1$ to this value we compare the FOM in different visible energy regions.
Between 3 and 12~${\rm MeV_{ee}}$ the FOM is calculated in 1~${\rm MeV_{ee}}$-wide bins, 
and thereafter in $3~{\rm MeV_{ee}}$ bins up to 18~${\rm MeV_{ee}}$.  
The results are shown in Figure~\ref{fig:fomenergyregion}. 
In general, the FOM values are larger than those of Figure~\ref{fig:fomintegraltime}
because the neutron peak narrows when specific energy ranges are considered, 
as can been seen by comparing, for instance, Figure~\ref{fig:psd1dplot} and Figure~\ref{fig:psdevaluate}. 
Note that the FOM values between 5 and 10~${\rm MeV_{ee}}$ are almost flat and higher than the 
other ranges. 
This range is the most important for the measurement of gamma-rays
emitted from the excited nuclei and the large FOM indicates that 
neutron backgrounds can be effectively rejected with the CsI(Tl) scintillator.
At lower energy, the FOM degrades due to the large spread in PSD parameters values 
of gamma-ray events.  
Above 10~${\rm MeV_{ee}}$ the FOM becomes smaller due to the tendency of 
neutron PSD parameters to become closer to the gamma-ray case, as seen in Figure~\ref{fig:psdplots}.

%///////////////////////////////////////////////////////////////////////////////////////

%\clearpage
%\input{conclusion}
\section{Conclusion}
\label{sec:conlusion}
In this paper we have demonstrated that neutron and gamma-ray discrimination is possible
using pulse shape information from CsI(Tl) scintillators.
Waveform data from both gamma-ray and fast neutron exposures of a single crystal at the CYRIC facility
indicate that the two populations can be clearly seen using the ratio of early to late
accumulated charge.
We have optimized the performance of the particle discrimination over the
relative lengths of these time periods using a standard figure of merit.
With the optimized charge integration regions we find that the
FOM takes on its highest values in the region between 5 and 10 ${\rm MeV_{ee}}$.
This region is of particular interest to the study of nuclear de-excitation gamma-rays,
and our measurements indicate that CsI(Tl) scintillators can effectively reject neutron
backgrounds to such measurements.

%\newpage
\section*{Acknowledgments}

This experiment was carried out in cooperation with the JSNS$^2$ group
at Tohoku University's Research Center for Neutrino Science.
We thank F. Suekane, H. Furuta, and Y. Hino for their assistance in conducting the experiment.
We also thank the CYRIC accelerator group for supplying the beam.
This work was supported by JSPS KAKENHI Grant Numbers 17J06141 and 26400292, and
Grant-in-Aid for Scientific Research on Innovative Areas
titled ``Unification and Development of the Neutrino Science Frontier"
(MEXT KAKENHI 25105002).

%% The Appendices part is started with the command \appendix;
%% appendix sections are then done as normal sections
%% \appendix

%% \section{}
%% \label{}

%% References
%%
%% Following citation commands can be used in the body text:
%% Usage of \cite is as follows:
%%   \cite{key}         ==>>  [#]
%%   \cite[chap. 2]{key} ==>> [#, chap. 2]
%%

%% References with bibTeX database:

%\bibliographystyle{elsarticle-num}
%\bibliography{<your-bib-database>}

\begin{thebibliography}{100}

%% \bibitem must have the following form:
%%   \bibitem{key}...
%%
 
%\bibitem{ankowski} Arthur M. Ankowski $et \ al.$, Phys, Rev. Lett. 108, 052505, 2012
%\bibitem{skdetector} S. Fukuda $et \ al.$, Nucl. Instr. and Meth. in Physics Research A, 
%                     501, 2003
%\bibitem{sksrn123} K. Bays $et \ al.$ (Super-Kamiokande Collaboration), Phys. Rev. D 85, 
%                   052007, 2012
%\bibitem{sksrn4} H. Zhang $et \ al.$ (Super-Kamiokande Collaboration), Astropart. Phys. 60, 
%                 41, 2015
%\bibitem{t2kdm} P. deNiverville $et \ al.$, Phys. Rev. D 86, 035022, 2012
\bibitem{leo} W. R. Leo, Springer-Verlag (1993)
\bibitem{knoll} G. F. Knoll, Jhon Wiley \& Sons (2001)
\bibitem{iwamoto} Y. Iwamoto $et \ al.$, Nucl. Instr. Meth. Phys. Res. A 
                  804 (2015)
\bibitem{wu} S. C. Wu $et \ al.$, Nucl. Instr. Meth. Phys. Res. A 
             523 (2004)
\bibitem{dcunha} M. F. D'Cunha $et \ al.$, Nucl. Instr. Meth. Phys. Res. A 
                 95 (1971)
%\bibitem{kamanin} D. V. Kamanin $et \ al.$, Nucl. Instr. and Meth. in Physics Research A 
%                  413, 1998
%\bibitem{stracener} D. W. Stracener $et \ al.$, Nucl. Instr. and Meth. in Physics Research A 
%                    294, 1990
%\bibitem{benrachi} F. Benrachi $et \ al.$, Nucl. Instr. and Meth. in Physics Research A 
%                   281, 1989
\bibitem{alarja} J. Alarja $et \ al.$, Nucl. Instr. Meth. Phys. Res. A 
                 242 (1986)
\bibitem{alderighi} M. Alderighi $et \ al.$, Nucl. Instr. Meth. Phys. Res. A 
                    489 (2002)
\bibitem{huang} K. Huang, JPS Conf. Proc. 12, 010052 (2016)
%\bibitem{t2kexp} K. Abe $et \ al.$ (T2K Collaboration), Nucl. Instr. and Meth.
%                 in Physics Research A 659, 2011
\bibitem{t2kncqe} K. Abe $et \ al.$ (T2K Collaboration), Phys. Rev. D 90, 072012 (2014)
\bibitem{orihara} H. Orihara $et \ al.$, Nucl. Instr. Meth. Phys. Res. A 
                  188 (1981)
\bibitem{soderstrom} P. -A. S\"{o}derstr\"{o}m $et \ al.$, 
                     Nucl. Instr. Meth. Phys. Res. A 594 (2008)

\end{thebibliography}

%% Authors are advised to submit their bibtex database files. They are
%% requested to list a bibtex style file in the manuscript if they do
%% not want to use elsarticle-num.bst.

%% References without bibTeX database:

%% \bibitem must have the following form:
%%   \bibitem{key}...
%%

\end{document}